\documentclass{article}
\usepackage{amssymb}
\begin{document}
\title{Symmetries of the Hamiltonian operator and constants of motion}

\author{G.F.\ Torres del Castillo \\ Departamento de F\'isica Matem\'atica, Instituto de Ciencias \\
Universidad Aut\'onoma de Puebla, 72570 Puebla, Pue., M\'exico \\[2ex]
J.E.\ Herrera Flores \\ Facultad de Ciencias F\'isico Matem\'aticas \\ Universidad Aut\'onoma de Puebla, 72570 Puebla, Pue., M\'exico}

\maketitle

\begin{abstract}
It is shown that, in the framework of non-relativistic quantum mechanics, any conserved Hermitian operator (which may depend explicitly on the time) is the generator of a one-parameter group of unitary symmetries of the Hamiltonian and that, conversely, any one-parameter family of unitary symmetries of the Hamiltonian is generated by a conserved Hermitian operator.
\end{abstract}


\section{Introduction}
The existence of a relationship between continuous symmetries and conservation laws is known in various areas of physics such as classical mechanics, field theory and quantum mechanics. The most widely known version of this relationship appears in the Lagrangian formulation, but its most complete version is found in the Hamiltonian formulation of classical mechanics where any constant of motion, without exception, is associated with a group of canonical transformations that leave the Hamiltonian invariant.

Owing to the similarities between the standard formulation of the non-relativistic quantum mechanics and the Hamiltonian formulation of classical mechanics, one can expect a relationship between an arbitrary constant of motion (that is, a Hermitian operator, $A$, satisfying the condition
\begin{equation}
{\rm i} \hbar \frac{\partial A}{\partial t} + [A, H] = 0, \label{cons}
\end{equation}
where $H$ is the Hamiltonian operator of the system) and some group of transformations that leave the Hamiltonian invariant.

The constants of motion considered in most of the textbooks on quantum mechanics do not depend explicitly on the time (see, e.g., Refs.\ \cite{AD,GM,JS,EM,NZ,SN,JP}). In that case, Eq.\ (\ref{cons}) reduces to $[A, H] = 0$, which implies that
\begin{equation}
\exp ({\rm i} sA/\hbar) \, H \exp(- {\rm i} sA/\hbar) = H, \label{inv}
\end{equation}
for all $s \in \mathbb{R}$; that is, $H$ is invariant under the one-parameter group of transformations generated by $A$ (the transformation $\exp(- {\rm i} sA/\hbar)$ is unitary if $A$ is Hermitian). Conversely, taking the derivative with respect to $s$, at $s = 0$, of both sides of Eq.\ (\ref{inv}) one finds that $A$ commutes with $H$ and, therefore, if $A$ does not depend on the time, then $A$ is conserved. It should be pointed out that Refs.\ \cite{GY,SW} also consider the Galilean transformations, which are related to a constant of motion that depends explicitly on the time (see Section 3.1, below).

The aim of this paper is to show that, with an appropriate definition of the invariance of a Hamiltonian [that generalizes Eq.\ (\ref{inv})], any conserved operator is associated with a group of unitary transformations that leave the Hamiltonian invariant, and to give illustrative examples of this relationship. In Section 2 we give the definition of the invariance of a Hamiltonian under a unitary transformation that may depend on the time and then we demonstrate the main results of this paper, establishing the connection between conserved operators and one-parameter families of unitary transformations that leave the Hamiltonian invariant. Section 3 contains several examples related to constants of motion that depend explicitly on the time.
	
\section{Invariance of a Hamiltonian under a unitary transformation}
We shall say that the unitary operator $U$ is a symmetry of the Hamiltonian $H$ (or that $H$ is invariant under $U$) if
\begin{equation}
U^{-1} H U = H + {\rm i} \hbar \, U^{-1} \frac{\partial U}{\partial t}, \label{invg}
\end{equation}
so that if $\psi$ is a solution of the Schr\"{o}dinger equation, then $U \psi$ is also a solution
\begin{eqnarray*}
{\rm i} \hbar \frac{\partial}{\partial t} (U \psi) - H (U \psi) & = & {\rm i} \hbar \left( \frac{\partial U}{\partial t} \right) \psi + U {\rm i} \hbar \frac{\partial \psi}{\partial t} - HU \psi \\
& = & {\rm i} \hbar \left( \frac{\partial U}{\partial t} \right) \psi + U H \psi - HU \psi \\
& = & U \left( {\rm i} \hbar \, U^{-1} \frac{\partial U}{\partial t} + H - U^{-1} H U \right) \psi \\
& = & 0.
\end{eqnarray*}
Note that, when $U$ does not depend on the time, condition (\ref{invg}) reduces to $H U = U H$ [cf.\ Eq.\ (\ref{inv})].

\noindent {\bf Proposition.} If $H$ is invariant under a one-parameter family of unitary transformations, $U_{s}$, then, assuming that $U_{0}$ is the identity operator,
\begin{equation}
A \equiv {\rm i} \hbar \left( U_{s}{}^{-1} \left. \frac{\partial U_{s}}{\partial s} \right) \right|_{s = 0} \label{gen}
\end{equation}
is a constant of motion. (Note that the operators $U_{s}$ need not form a group.)

\noindent {\it Proof.} If $H$ is invariant under the transformations $U_{s}$ then, taking the derivatives with respect to $s$ of both sides of
\[
U_{s}{}^{-1} \, H U_{s} = H + {\rm i} \hbar \, U_{s}{}^{-1} \frac{\partial}{\partial t} U_{s}
\]
at $s = 0$ [see Eq.\ (\ref{invg})] we obtain
\[
- \frac{1}{{\rm i} \hbar} A H + H \frac{1}{{\rm i} \hbar} A = {\rm i} \hbar \frac{\partial}{\partial t} \left( \frac{1}{{\rm i} \hbar} A \right),
\]
which means that $A$ is conserved [see Eq.\ (\ref{cons})].

Conversely, any constant of motion generates a one-parameter group of symmetries of $H$.

\noindent {\bf Proposition.} If the Hermitian operator $A$ is conserved, then $H$ is invariant under the one-parameter group of unitary transformations $U_{s} = \exp (- {\rm i} s A/\hbar)$.

\noindent {\it Proof.} If $A$ is conserved, so it is any power of $A$ and, therefore, $U_{s} = \exp (- {\rm i} s A/\hbar)$ also satisfies Eq.\ (\ref{cons}), for any value of the parameter $s$, which amounts to say that $U_{s}$ satisfies Eq.\ (\ref{invg}).

\section{Examples}
In this section we give some examples related to the Propositions proved above. In these examples, the symmetry transformations and the corresponding constants of motion depend explicitly on the time.

\subsection{Galilean transformations}
The action of a Galilean transformation on the position and momentum operators of a particle of mass $m$ is given by
\begin{equation}
U_{v}{}^{-1} x \, U_{v} = x - vt, \qquad U_{v}{}^{-1} p \, U_{v} = p - mv \label{gt}
\end{equation}
where the parameter $s$ has been replaced by $v$, which represents the relative velocity between two inertial frames. Differentiating Eqs.\ (\ref{gt}) with respect to $v$ at $v = 0$, we obtain
\[
\frac{1}{{\rm i} \hbar} [A, x] = - t, \qquad \frac{1}{{\rm i} \hbar} [A, p] = - m
\]
[see Eq.\ (\ref{gen})], which imply that
\begin{equation}
A = mx - pt + f(t), \label{gg}
\end{equation}
where $f(t)$ is a real-valued function of $t$ only [cf.\ Ref.\ \cite{GY}, Sec.\ 7.3, Eq.\ (61)]. The function $f$ {\em is not}\/ determined by Eqs.\ (\ref{gt}), and can be chosen in such a way that the invariance condition (\ref{invg}) is satisfied, once the Hamiltonian is chosen. For instance, in the case of a free particle, one readily finds that $A$ is conserved if we take $f = 0$.

On the other hand, if we look for the most general Hamiltonian of the form
\begin{equation}
H = \frac{p^{2}}{2m} + V(x,t), \label{ham}
\end{equation}
that is invariant under the Galilean transformations, substituting (\ref{gg}) and (\ref{ham}) into Eq.\ (\ref{cons}), we find that
\[
V(x, t) = - \frac{1}{t} \frac{{\rm d} f}{{\rm d} t} x,
\]
which corresponds to a, possibly time-dependent, uniform field. For instance, for a constant, uniform field, $V = - eEx$, where $e$ and $E$ are constants,
\[
f = {\textstyle \frac{1}{2}} e E t^{2}
\]
and, therefore,
\[
A = mx - pt + {\textstyle \frac{1}{2}} e E t^{2}
\]
is, in this case, the (conserved) generator of the Galilean transformations.

\subsection{The symmetry transformations generated by a given constant of motion}
A straightforward computation shows that the operator
\begin{equation}
A = \frac{1}{2} (xp + px) + \frac{3eEt^{2} p}{2m} - \frac{tp^{2}}{m} - eEtx - \frac{e^{2} E^{2} t^{3}}{2m} \label{c}
\end{equation}
is a constant of motion if the Hamiltonian is chosen as
\[
H = \frac{p^{2}}{2m} - eE x,
\]
where $e$ and $E$ are constants. We can compute the action of $U_{s} = \exp (- {\rm i} s A/\hbar)$ on the operators $x$ and $p$ with the aid of the well-known formula
\[
{\rm e}^{X} Y {\rm e}^{- X} = Y + [X, Y] + \frac{1}{2!} [X, [X, Y]] + \frac{1}{3!} [X, [X, [X, Y]]] + \cdots,
\]
which yields, e.g.,
\[
U_{s}{}^{-1} p \, U_{s} = p + \frac{{\rm i} s}{\hbar} [A, p] + \frac{1}{2!} \left( \frac{{\rm i} s}{\hbar} \right)^{2} [A, [A, p]] + \frac{1}{3!} \left( \frac{{\rm i} s}{\hbar} \right)^{3} [A, [A, [A, p]]] + \cdots.
\]
By means of a straightforward computation one obtains
\begin{eqnarray*}
\mbox{} [A, p] & = & {\rm i} \hbar (p - eEt), \\
\mbox{} [A, [A, p]] & = & ({\rm i} \hbar)^{2} (p - eEt), \\
\mbox{} [A, [A, [A, p]]] & = & ({\rm i} \hbar)^{3} (p - eEt),
\end{eqnarray*}
and so on. Hence,
\[
U_{s}{}^{-1} p \, U_{s} = p + ({\rm e}^{-s} - 1) (p - eEt).
\]
In a similar manner, we obtain
\[
U_{s}{}^{-1} x \, U_{s} = x + \left( x + \frac{eEt^{2}}{2m} - \frac{tp}{m} \right) ({\rm e}^{s} - 1) + \left( \frac{tp}{m} - \frac{eEt^{2}}{m} \right) ({\rm e}^{-s} - 1).
\]
According to the second Proposition of Section 2, these transformations must be a symmetry group of the Hamiltonian.

\subsection{An example involving spin}
In the case of the spin of an electron in a static, uniform magnetic field in the $z$-direction, the Hamiltonian is taken as (see, e.g., Ref.\ \cite{SN}, sec.\ 2.1)
\[
H = \omega S_{z},
\]
where $\omega \equiv |e| B/mc$, $e$ is the electric charge of the electron, $m$ is its mass, and $S_{z}$ is the $z$-component of its spin. Then, the operator
\[
A = \cos \omega t S_{x} + \sin \omega t S_{y}
\]
is a constant of motion and we can find the explicit expression of $U_{s} = \exp (- {\rm i} sA/\hbar)$, namely,
\[
U_{s} = \cos (s/2) - {\rm i} \sin (s/2) \; 2A/\hbar.
\]

The state ${\rm e}^{- {\rm i} \omega t/2} \, |+\rangle$, where $|+\rangle$ is an eigenket of $S_{z}$ with eigenvalue $\hbar/2$, is a ({\em stationary}) solution of the Schr\"odinger equation and, therefore, $U_{s} \, {\rm e}^{- {\rm i} \omega t/2} \, |+\rangle$ must be also a solution, for any value of $s$. We obtain
\[
U_{s} \, {\rm e}^{- {\rm i} \omega t/2} \, |+\rangle \,  = \cos (s/2) \, {\rm e}^{- {\rm i} \omega t/2} \, |+\rangle - {\rm i} \sin (s/2) \, {\rm e}^{{\rm i} \omega t/2} \, |-\rangle,
\]
which is indeed a solution of the Schr\"odinger equation for all values of $s$. Only for $s = 0, \pm \pi, \pm 2 \pi, \ldots \,$, one obtains a stationary state.

\section{Final remarks}
It is noteworthy that the proofs of the Propositions presented in Section 2 are simpler than their analogs in classical mechanics. Part of this simplification comes from the fact that, in the formalism of quantum mechanics, both the conserved quantities and the transformations correspond to operators. In fact, the form of Eq.\ (\ref{cons}) coincides with that of Eq.\ (\ref{invg}) for an invertible operator.

As pointed out in Section 2, the action of a symmetry operator $U$ on a solution of the Schr\"{o}dinger equation yields another solution of the same equation, but when $U$ depends explicitly on the time the image of a stationary state will not always be a stationary state.

\section*{Acknowledgement}
One of the authors (J.E.H.F.) wishes to thank the Vicerrector\'{\i}a de Investigaci\'on y Estudios de Posgrado of the Benem\'erita Universidad Aut\'onoma de Puebla for financial support.

\end{document}